# Vortex Transport Entropy in cuprate superconductors and Boltzmann Constant

R. P. Huebener[*], Physikalisches Institut, Universität Tübingen, 72076 Tübingen, Germany

and

H.-C. Ri[†], Department of Physics, Kyungpook National University, Daegu 41566, South Korea

Abstract

The vortex transport entropy in the mixed state of the cuprate superconductors $YBa_2Cu_3O_{7-\delta}$ and $Bi_2Sr_2CaCu_2O_{8+x}$ at the maximum Nernst or Ettingshausen signal below $T_c$ is discussed. The vortex transport entropy per $CuO_2$ double layer in the cuprate epitaxial films is found to be close to the Boltzmann constant.

## 1. Introduction

The Nernst effect and Ettingshausen effect in the mixed state of superconductors have been studied since many years [1]. In the following we focus on the vortex transport entropy $S_\varphi$ of a magnetic flux quantum associated with thermal diffusion of the vortices or with the heat current density carried by the moving vortices. Thermal diffusion is caused by the thermal force

$$\mathbf{f}_{th} = -S_\varphi \nabla T \qquad (1)$$

acting on the vortices in a temperature gradient. The heat current density

$$\mathbf{U} = nTS_\varphi \mathbf{v}_\varphi \qquad (2)$$

arises due to the Lorentz force

$$\mathbf{f}_L = \mathbf{j} \times \boldsymbol{\varphi}_0 \qquad (3)$$

of an applied electric current of density **j** acting on the vortex lines. In (2) – (3) $n$ is the areal density of the magnetic flux quanta, $\mathbf{v}_\varphi$ is the vortex velocity, and $\boldsymbol{\varphi}_0$ the magnetic flux quantum. In (1) – (3) the forces and $S_\varphi$ are taken per unit length of magnetic flux line.

In the case of the Nernst effect the thermal force (1) drives the vortex lattice down the temperature gradient from the hot to the cold side. If the negative temperature gradient and the vortex motion is oriented in $x$-direction and the magnetic flux density **B** in $z$-direction, the flux-flow electric field $E_y$ is directed in $y$-direction. From the balance of forces one finds the transport entropy [2]

$$S_\varphi = (E_y/\nabla_x T)\,(\varphi_0/\rho) \tag{4}$$

Here $\rho$ is the flux-flow resistivity.

In the case of the Ettingshausen effect an applied electric current along the $x$-direction causes flux motion and a heat current density

$$U_y = nTS_\varphi v_{\varphi y} \tag{5}$$

in $y$-direction. From the force balance one finds [1]

$$S_\varphi = [(\partial T/\partial y)/E_x][\kappa \varphi_0/T] \tag{6}$$

where $\kappa$ denotes the heat conductivity.

Recently, C. W. Rischau et al. reported [3], that the vortex transport entropy in the mixed state of two-dimensional heterostructures at the maximum Nernst signal below $T_c$ is close to Boltzmann constant. This means that the pancake vortices must be treated in thermodynamics as individual particles, similarly as in the case of atoms and molecules, (where we ignore electrons, protons, neutrons, quarks, gluons, etc., since these refer to much higher energies). Similar ideas were discussed previously by K. A. Moler et al. [4].

From these results the question arises, whether the $CuO_2$ planes (or the $CuO_2$ double planes) in the cuprate superconductors also show the behavior found in Ref. [3] in the two-dimensional heterostructures. Then, in the case of the cuprates, the vortex transport entropy $S_\varphi$, given per unit length of magnetic flux line, must be calculated per lattice constant of the material in the direction of the applied magnetic field. This yields the contribution to $S_\varphi$ from the $CuO_2$ planes passing through the unit cell. In the following we present the results of such a calculation in the case of two cuprate superconductors.

## 2. Discussion

We calculate $S_\varphi$ from the maximum Nernst or Ettingshausen signal of the experimental data using equations (4) and (6). Our data are taken from experiments performed with the cuprate superconductors $YBa_2Cu_3O_{7-\delta}$ and $Bi_2Sr_2CaCu_2O_{8+x}$. Finally, we multiply $S_\varphi$ (given per unit length of a magnetic flux line) with the lattice constant of the material in the direction of the applied magnetic field, in order to obtain the contribution to $S_\varphi$ from the individual $CuO_2$ planes. Since the magnetic field had been applied along the $c$-direction and perpendicular to the $CuO_2$ planes, we used $c = 11.9 \times 10^{-10}$ m in the case of $YBa_2Cu_3O_{7-\delta}$ and $c = 30.7 \times 10^{-10}$ m in the case of $Bi_2Sr_2CaCu_2O_{8+x}$.

In Table 1 we show the results of these calculations of $S_\varphi c$. The data in the case of the epitaxial films of $YBa_2Cu_3O_{7-\delta}$ and of $Bi_2Sr_2CaCu_2O_{8+x}$ (Nernst effect) are taken from H.-C. Ri et al. [5] and for the $YBa_2Cu_3O_{7-\delta}$ single crystal (Ettingshausen effect) from T. T. M. Palstra et al. [6].

Table 1: Calculated values of $S_\varphi c$ from the experimental data of H.-C. Ri et al. [5] on epitaxial films of $YBa_2Cu_3O_{7-\delta}$ and of $Bi_2Sr_2CaCu_2O_{8+x}$ and of T. T. M. Palstra et al. [6] on single crystals of $YBa_2Cu_3O_{7-\delta}$.

| Material | Experiment | B(T) | $S_\varphi c$ |
|---|---|---|---|
| YBCO epitaxial film | Nernst effect | 4 | 0.69 x $10^{-23}$ J/K |
| " | " | 10 | 1.31 x $10^{-23}$ J/K |
| " | " | 12 | 1.38 x $10^{-23}$ J/K |
| YBCO single crystal | Ettingshausen effect | 5 | 4.88 x $10^{-23}$ J/K |
| " | " | 10 | 5.47 x $10^{-23}$ J/K |
| BSCCO epitax. Film | Nernst effect | 4 | 4.26 x $10^{-23}$ J/K |
| " | " | 10 | 5.12 x $10^{-23}$ J/K |
| " | " | 12 | 5.12 x $10^{-23}$ J/K |

As we see from Table 1, in the case of the Nernst effect in the epitaxial film of $YBa_2Cu_3O_{7-\delta}$ the calculated values of $S_\varphi c$, indeed, are close to Boltzmann constant $k_B = 1.38 \times 10^{-23}$ J/K. In this case one $CuO_2$ double layer is passing through the unit cell. Why the values of $S_\varphi c$ in the case of the Ettingshausen effect in an $YBa_2Cu_3O_7$ single crystal are higher than $k_B$ by a factor of $3-4$, different from the epitaxial film, is unclear at present. In the case of the Nernst effect in the epitaxial $Bi_2Sr_2CaCu_2O_{8+x}$ film we are dealing with two $CuO_2$ double layers passing through the unit cell. Hence, in this case the vortex transport entropy per unit cell is expected to be higher than in $YBa_2Cu_3O_{7-\delta}$ by a factor of two.

## 3. Conclusion

We find that the vortex transport entropy per $CuO_2$ double layer in epitaxial $YBa_2Cu_3O_{7-\delta}$ and $Bi_2Sr_2CaCu_2O_{8+x}$ films is of the order of $k_B$. We note that these questions about the entropy of the moving vortex lattice deserve further experimental and theoretical attention.

## Acknowledgment

One of the authors (R P H) is grateful to A. Kapitulnik for pointing out to him Ref. [4].

---

[*]prof.huebener@uni-tuebingen.de

[†]hcri@knu.ac.kr